\documentclass[pra,twocolumn,tightenlines,showpacs,nofootinbib]{revtex4}
\usepackage{bm,dcolumn,amsmath,graphicx,amsfonts,amssymb}

\begin{document}
\title{Calculation of parity nonconservation in neutral ytterbium}

\author{V. A. Dzuba}
\affiliation{School of Physics, University of New South Wales,
Sydney, NSW 2052, Australia}
\author{V. V. Flambaum}
\affiliation{School of Physics, University of New South Wales,
Sydney, NSW 2052, Australia}

\date{ \today }

\begin{abstract}
We use configuration interaction and many-body perturbation theory
techniques to calculate spin-independent and spin-dependent parts of
the parity nonconserving amplitudes of the transitions between the
$6s^2 \ ^1$S$_0$ ground state and the $6s5d \ ^3$D$_1$  excited state
of $^{171}$Yb and $^{173}$Yb. The results 
are presented in a form convenient for extracting
spin-dependent 
interaction constants (such as, e.g., anapole moment) from the
measurements. 
\end{abstract}
\maketitle

\section{Introduction}

The use of atomic ytterbium to study parity nonconservation (PNC) in
atoms was first suggested by DeMille~\cite{DeMille}. The measurements
are in progress at Berkeley and afters years of hard
work~\cite{Bowers,Kimball,Stalnaker02,Stalnaker06} the first results
of PNC measurements are finally reported~\cite{BudkerYbPNC}. As it
was expected the PNC in ytterbium is strongly enhanced, being two orders
of magnitude larger than in cesium~\cite{Wood}. The cesium PNC
experiment together with its interpretation~ \cite{Breit,QED,Cs-cor} in
terms of nuclear weak 
charge provides the best current atomic test of the standard model
(see, also a review~\cite{Ginges}). It is also the only measurement
of the nuclear  {\em anapole moment} which is produced by the PNC nuclear
 forces \cite{anapole}.
 The extraction of the weak
nuclear charge from the PNC measurements relies on atomic
calculations. The interpretation of the PNC measurements in
ytterbium similar to what was done for cesium is not possible due to
limitations of atomic theory. Ytterbium has complicated electron
structure and calculations for it on the same level of accuracy as for
cesium are not possible now and in foreseen future. The aims of the
PNC measurements in Yb are different~\cite{BudkerYbPNC}: (i) to study
the ratio of the PNC amplitudes for different isotopes, and (ii) to
measure nuclear spin-dependent PNC effects, such as the effect of
nuclear anapole moment. The study of the PNC for a chain of isotopes
does not require atomic calculations and can deliver useful
information about either neutron distribution or new physics beyond
standard model (see, e.g.~\cite{DFK86,Fortson,DP02,BDF09}). The
extraction of the anapole moment from the measurements does require
atomic calculations, however, high theoretical accuracy is not critical here.

Ytterbium is a very good candidate for both types of the experimental
studies. It has seven stable isotopes with large difference in neutron
numbers $\Delta N_{max} = 8$. Two of the isotopes, $^{171}$Yb and
$^{173}$Yb, have non-zero nuclear spin provided by valence
neutron. This is especially interesting since 
it allows one to measure the strength of the neutron-nucleus PNC potential
\cite{anapole} (the anapole moment has been measured only for the $^{133}$Cs
 nucleus which has valence proton).

Calculations of the spin-independent PNC in ytterbium were performed in
Refs.~\cite{DeMille,Porsev95,Das97}. Calculations of the
spin-dependent PNC were reported in \cite{Singh99,Porsev00}.
The results of~\cite{Singh99,Porsev00} for
the spin-dependent PNC amplitudes are presented as tables of reduced
matrix elements of the spin-dependent weak interaction for different
hyperfine transitions. This, in our view, leads to some difficulties
in interpretation. Reduced matrix elements (RME) are very convenient for
intermediate calculations. However, presenting final results in a form
of RME may lead to confusion due to their unnatural
symmetry properties:
\begin{equation}
  \langle F_a,a || \hat H|| F_b,b \rangle = (-1)^{F_a-F_b} \langle
  F_b,b || \hat H|| F_a,a \rangle^*. \label{eq:sym}
\end{equation}
Here $F_a$ is the total momentum of the state $a$, asterisk means
complex conjugation which in the case of PNC amplitudes means the
change of sign. There is an apparent disagreement between the signs of
different RME in \cite{Singh99} and \cite{Porsev00}. The most likely
explanation for this in our view is that the authors of \cite{Singh99}
and \cite{Porsev00} presented different RME, say $\langle a||\hat H||
b \rangle$ in \cite{Singh99} and $\langle b||\hat H||a \rangle$ in 
\cite{Porsev00}. At least all sign differences follow strictly the
rule (\ref{eq:sym}). 

 Strictly speaking, the sign of an amplitude is not defined
 (since a wave function may be multiplied by an arbitrary phase factor),
 only the ratio of two amplitudes between the same states has definite sign.
  Neither of the work~\cite{Singh99,Porsev00} provides a link between the
spin-dependent (SD) PNC amplitudes and spin-independent (SI) PNC amplitudes
calculated earlier in \cite{Porsev95,Das97}. This means that it is
hard to say whether the spin-dependent effects
increase or decrease a particular PNC amplitude. In other words, the sign
of the spin-dependent interaction constants, such as the anapole
moment, cannot be extracted from the measurements when using the
calculations of \cite{Singh99} or \cite{Porsev00}
 and no additional assumptions (note that the apparent disagreement between
 the signs of the amplitudes in \cite{Singh99} and \cite{Porsev00} shows that
any guesswork about the relative signs of the SI and SD amplitudes is
unreliable). 

To avoid this problem, in present paper 
 both spin-independent and spin-dependent PNC amplitudes are calculated
simultaneously using the same procedure and the same wave
functions. In this approach the relative sign of the amplitudes is
fixed. This allows for unambiguous determination of the sign of the
spin-dependent contribution. The
constant of the spin-dependent interaction
can be expressed via the ratio of the two amplitudes. This
brings an extra advantage of more accurate interpretation of the
measurements. The accuracy of the calculations for the ratio of the PNC
amplitudes is higher than that for each of the amplitudes. This is
because the amplitudes are very similar in nature and most of the
theoretical uncertainty cancels out in the ratio.

\section{Theory}

Hamiltonian describing parity-nonconserving electron-nuclear
interaction can be written as a sum of spin-independent (SI) and
spin-dependent (SD) parts (we use atomic units: $\hbar = |e| = m_e = 1$):
\begin{eqnarray}
     H_{\rm PNC} &=& H_{\rm SI} + H_{\rm SD} \nonumber \\
      &=& \frac{G_F}{\sqrt{2}}                             
     \Bigl(-\frac{Q_W}{2} \gamma_5 + \frac{\varkappa}{I}
     {\bm \alpha} {\bm I} \Bigr) \rho({\bm r}),
\label{e1}
\end{eqnarray}
where  $G_F \approx 2.2225 \times 10^{-14}$ a.u. is the Fermi constant of
the weak interaction, $Q_W$ is the nuclear weak charge,
$\bm\alpha=\left(
\begin{array}
[c]{cc}%
0 & \bm\sigma\\
\bm\sigma & 0
\end{array}
\right)$ and $\gamma_5$ are the Dirac matrices, $\bm I$ is the
nuclear spin, and $\rho({\bf r})$ is the nuclear density normalized to 1.
The strength of the spin-dependent PNC interaction is proportional to
the dimensionless constant $\varkappa$ which is to be found from the
measurements. There are three major contributions to
$\varkappa$ arising from (i) electromagnetic interaction of atomic
electrons with nuclear {\em anapole moment}, (ii) electron-nucleus
spin-dependent weak interaction, and (iii) combined effect of
spin-independent weak interaction and magnetic hyperfine interaction
(see, e.g. ~\cite{Ginges}). In this work we do not distinguish
between different contributions to $\varkappa$ and present the results
in terms of total $\varkappa$ which is the sum of all possible
contributions. 

Within the standard model
the weak nuclear charge $Q_W$ is given by~\cite{PDG}
\begin{equation}
Q_W \approx -0.9877N + 0.0716Z.
\end{equation}
Here $N$ is the number of neutrons, $Z$ is the number of protons.

The PNC amplitude of an electric dipole transition between states of
the same parity $|i\rangle$ and $|f \rangle$ is equal to:
\begin{eqnarray}
   E1^{PNC}_{fi}  &=&  \sum_{n} \left[
\frac{\langle f | {\bm d} | n  \rangle
      \langle n | H_{\rm PNC} | i \rangle}{E_i - E_n}\right.
\nonumber \\
      &+&
\left.\frac{\langle f | H_{\rm PNC} | n  \rangle
      \langle n | d_q | i \rangle}{E_f - E_n} \right],
\label{eq:e2}
\end{eqnarray}
where ${\bm d} = -e\sum_i {\bm r_i}$ is the electric dipole operator,
  $|a \rangle \equiv |J_a F_a M_a \rangle$ and ${\bm F} = {\bm I}
+ {\bm J}$ is the total angular momentum. 

Applying the Wigner-Eckart theorem we can express the amplitudes via
reduced matrix elements
\begin{eqnarray}
  E1^{PNC}_{fi} &=&
      (-1)^{F_f-M_f} \left( \begin{array}{ccc}
                           F_f & 1 & F_i  \\
                          -M_f & q & M_i   \\
                           \end{array} \right) \nonumber \\
   &\times& \langle J_f F_f || d_{\rm PNC} || J_i F_i \rangle .
\end{eqnarray}
Detailed expressions for the reduced matrix elements of the SI and
SD PNC amplitudes can be found e.g. in Refs.~\cite{Porsev01} and
\cite{JSS03}. For the SI amplitude we have
\begin{eqnarray}
&&\langle J_f,F_f || d_{\rm SI} || J_i,F_i \rangle =
(-1)^{I+F_i+J_f+1}\nonumber \\ 
&& \times \sqrt{(2F_f+1)(2F_i+1)} 
\left\{ \begin{array}{ccc} J_i & J_f & 1 \\
                          F_f & F_i & I \\ 
                    \end{array} \right\}  \label{eq:si0}\\
&&  \times \sum_{n} \left[
\frac{\langle J_f || {\bm d} || n,J_n  \rangle
      \langle n,J_n || H_{\rm SI} || J_i \rangle}{E_i - E_n}\right.  \nonumber \\
&& + \left.\frac{\langle J_f || H_{\rm SI} || n,J_n  \rangle
      \langle n,J_n || {\bm d} || J_i \rangle}{E_f - E_n} \right] \nonumber \\
&& \equiv c(F_f,J_f,F_i,J_i) E^{\prime}_{fi}. \nonumber
\end{eqnarray}
Here $c(F_f,J_f,F_i,J_i)$ is the angular coefficient and the sum over $n$,
$E^{\prime}_{fi}$ does not depend on $F_f$ or $F_i$:
\begin{eqnarray}
  E^{\prime}  &=&  \sum_{n} \left[
\frac{\langle J_f || {\bm d} || n,J_n  \rangle
      \langle n,J_n || H_{\rm SI} || J_i \rangle}{E_i - E_n}\right. \label{eq:si} \\
&+&  \left.\frac{\langle J_f || H_{\rm SI} || n,J_n  \rangle
      \langle n,J_n || {\bm d} || J_i \rangle}{E_f - E_n} \right]. \nonumber 
\end{eqnarray}

For the SD PNC amplitude we have
\begin{eqnarray}
    && \langle J_f,F_f || d_{\rm SD} || J_i,F_i \rangle =
    \frac{G_F}{\sqrt{2}} \varkappa \nonumber \\
     &&\times  \sqrt{(I+1)(2I+1)(2F_i+1)(2F_f+1)/I}  \nonumber \\
    &&\times
     \sum_{n} \left[ (-1)^{J_f - J_i}
     \left\{ \begin{array}{ccc}
     J_n  &  J_i  &   1    \\
      I   &   I   &  F_i   \\                                  
     \end{array} \right\}
     \left\{ \begin{array}{ccc}
      J_n  &  J_f  &  1   \\
      F_f  &  F_i  &  I   \\
     \end{array} \right\} \right. \nonumber \\
  &&\times \frac{ \langle J_f || {\bm d} || n, J_n \rangle
     \langle n, J_n || {\bm \alpha}\rho || J_i \rangle }{E_n -
     E_i} \label{Eq:dsd}  \\
  &&+
     (-1)^{F_f - F_i}
     \left\{ \begin{array}{ccc}
     J_n  &  J_f  &   1    \\
      I   &   I   &  F_f   \\
     \end{array} \right\}
     \left\{ \begin{array}{ccc}
     J_n  &  J_i  &  1   \\
     F_i  &  F_f  &  I   \\
     \end{array} \right\} \nonumber \\
 &&\times
     \left. \frac{\langle J_f || {\bm \alpha}\rho ||n,J_n \rangle
            \langle n,J_n || {\bm d} ||J_i \rangle}{E_n - E_f}  \right].
\nonumber
\end{eqnarray}
In the case of the $^1$S$_0 - \rightarrow ^3$D$_1$ transition these
expressions can be significantly simplified. Substituting $F_i=I$,
$J_i=0$, $F_f =I,I \pm 1 \equiv F$, $J_f=1, J_n=1$ we have for the PNC
amplitudes ($z$-components) 
$E_{F_i,F_f}$ of the transitions between specific hfs states of
$^{171}$Yb ($I=1/2$)  
\begin{eqnarray}
  E_{\frac{1}{2},\frac{1}{2},z} &=& -\frac{1}{3}E^{\prime} Q_W -
  \sqrt{\frac{2}{27}} E^{\prime\prime}\varkappa, \label{1-2-1-2}\\ 
  E_{\frac{1}{2},\frac{3}{2},z} &=& \sqrt{\frac{2}{9}}E^{\prime} Q_W -
  \sqrt{\frac{1}{27}} E^{\prime\prime}\varkappa. \label{1-2-3-2} 
\end{eqnarray}
Similar expressions for $^{173}$Yb ($I=5/2$) are
\begin{eqnarray}
  E_{\frac{5}{2},\frac{3}{2},z} =&-& \sqrt{\frac{4}{45}}E^{\prime} Q_W 
 - \sqrt{\frac{98}{3375}} E^{\prime\prime}\varkappa, \label{5-2-3-2}\\ 
  E_{\frac{5}{2},\frac{5}{2},z} =&-& \sqrt{\frac{5}{21}}E^{\prime} Q_W 
 - \sqrt{\frac{2}{315}} E^{\prime\prime}\varkappa, \label{5-2-5-2}  \\
  E_{\frac{5}{2},\frac{7}{2},z} =&& \sqrt{\frac{2}{21}}E^{\prime} Q_W 
  - \sqrt{\frac{1}{63}} E^{\prime\prime}\varkappa \label{5-2-7-2}.
\end{eqnarray}
Here $E^{\prime\prime}$ is the part of the SD PNC amplitude which is
independent on $F_i$ and $F_f$:
\begin{eqnarray}
  E^{\prime\prime} &=& \frac{G_F}{\sqrt{2}}\sum_n \left[ \frac{
     \langle J_f || {\bm d} || n, J_n \rangle 
     \langle n, J_n || {\bm \alpha}\rho || J_i \rangle }{E_n - E_i} \right. \nonumber \\
  &-&   \left. \frac{\langle J_f || {\bm \alpha}\rho ||n,J_n \rangle
            \langle n,J_n || {\bm d} ||J_i \rangle}{E_n - E_f}  \right].
\label{eq:epp}
\end{eqnarray}
Note that if at least two PNC amplitudes are measured then the value
of $\varkappa$ can be expressed via the ratio
$E^{\prime\prime}/E^{\prime}$ of the calculated SD and SI PNC
amplitudes. This ratio is much less sensitive to numerical
uncertainties than each of the amplitudes. The amplitudes are very
similar. Therefore, a greater part
of the numerical uncertainty cancels out in the ratio.
For example, if amplitudes (\ref{1-2-1-2}) and  (\ref{1-2-3-2}) are
measured then
\begin{equation}
  \varkappa = \sqrt{3}\frac{\sqrt{2}R+1}{S(R-\sqrt{2})},
\label{eq:kappa}
\end{equation}
where $R= E_{\frac{1}{2},\frac{1}{2},z}/E_{\frac{1}{2},\frac{3}{2},z}$
and $S=E^{\prime\prime}/(E^{\prime}Q_W)$.
 The ratio of theoretical
 amplitudes $E^{\prime\prime}/E^{\prime}$  is significantly more stable in the
calculations than each of the amplitudes.

\section{Calculations}

We consider ytterbium as an atom with two valence electrons above
closed shells and use the combination of the configuration interaction
and many-body perturbation theory (CI+MBPT,~\cite{DzuFlaKoz96b}) to
perform the calculations. The calculations are very similar to our
previous calculations of ytterbium polarizabilities~\cite{DD10}. Below
we briefly describe the procedure emphasizing some minor differences.

\subsection{CI+MBPT method}

The effective CI+MBPT Hamiltonian for two valence electrons has the form
\begin{equation}
  \hat H^{\rm eff} = \hat h_1(r_1) + \hat h_1(r_2) + \hat h_2(r_1,r_2),
\label{Heff}
\end{equation}
where $\hat h_1$ is the single-electron part of the relativistic Hamiltonian
\begin{equation}
  \hat h_1 = c \mathbf{\hat{\alpha}} \mathbf{p} + (\hat{\beta}-1)m_e c^2-\frac{Ze^2}{r}
  + V^{N-2} + \hat \Sigma_1,
\label{h1}
\end{equation}
and $\hat h_2$ is the two-electron part of the Hamiltonian
\begin{equation}
  \hat h_2(r_1,r_2) = \frac{e^2}{|\mathbf{r}_1 - \mathbf{r}_2|} + \hat
  \Sigma_2(r_1,r_2).
\label{h2}
\end{equation}
In these equations, $\mathbf{\hat{\alpha}}$ and $\hat{\beta}$ are the
conventional Dirac matrices, 
$V^{N-2}$ is the Dirac-Hartree-Fock (DHF) potential of the closed-shell atomic
core ($N-2=68,Z=70$), and $\hat \Sigma$ is the correlation operator. It
represents terms in the Hamiltonian arising due to virtual excitations from
atomic core (see Ref.~\cite{DzuFlaKoz96b,DzuJoh98} for details).
$\hat \Sigma \equiv 0$ corresponds to the standard CI method.
$\hat \Sigma_1$ is a single-electron operator. It represents a
correlation interaction (core-polarization) of a particular valence
electron with the atomic 
core. $\hat \Sigma_2$ is a two-electron operator. It represents
screening of the Coulomb interaction between the two valence electrons
by the core 
electrons. We calculate $\hat \Sigma$ in the second order of the
MBPT. We use a  B-spline technique~\cite{JohSap86} to construct a
complete set of single-electron orbitals. We use 40 B-splines in a
cavity of radius $R=40\, a_B$ and calculate the eigenstates of the $V^{N-2}$
DHF Hamiltonian up to the maximum value of the angular
momentum $l_{max}=5$. The same basis is used in computing $\hat
\Sigma$ and in constructing the two-electron states for the valence
electrons. 40 out of 60 lowest-energy states for every $l$ up to $l_{max}=5$
are used to calculate $\hat \Sigma$ and 16 lowest states above the
core are used for every $l$ up to $l_{max}=4$ to construct the
two-electron states.

The two-electron valence states are found by solving the eigenvalue
problem,
\begin{equation}
  \hat H^{\rm eff} \Psi_v = E_v \Psi_v \, ,
\label{CI}
\end{equation}
using the standard CI techniques. Calculated and experimental energies of
a few lowest-energy states of Yb can be found in Ref.~\cite{DD10}.
The pure {\em ab initio} energies are already close to the
experimental values. However, for improving the accuracy further, we
re-scale the correlation operator $\hat \Sigma_1$ by replacing $\hat
\Sigma_1$ in the effective Hamiltonian (\ref{Heff}) in each partial
wave $s, p_{1/2}, p_{3/2} , \ldots$  by $f_a\hat \Sigma_{1}$. The
rescaling factors are $f_s=0.875$, $f_p=1.268$, $f_d=0.935$, and
$f_f=1$. These values are chosen to fit the experimental spectrum of
Yb. Some differences in scaling parameters compare to what was used in
Ref.~\cite{DD10} is due to the fact that in present work we have
fitted exactly the energy of the $^1$P$^o_1$ state while in
~\cite{DD10} we fitted the energy of the $^3$P$^o_1$ state.

\subsection{Dalgarno-Lewis and RPA methods}

Matrix elements are found with the random-phase approximation (RPA)
\cite{DzuGin06,DzuFla07}
\begin{equation}
  E1_{vw} = \langle \Psi_v || \hat f + \delta V^{N-2} || \Psi_w
  \rangle, \label{E1} 
\end{equation}
where $\delta V^{N-2}$ is the correction to the core potential due to core
polarization by an external field $\hat f$. In present calculations
$\hat f$ represents either external electric field, SI weak
interaction or SD PNC interaction.

Computing PNC requires summing over a complete set of two-electron
states (see, e.g. Eq.~(\ref{eq:e2})).
We use the Dalgarno-Lewis method~\cite{DalLew55} for the summation.
In this method, a correction $\delta \Psi_v$ to the two-electron wave
function of the state $v$ is introduced and the amplitude is reduced to
\begin{equation}
  A_{vw} = \langle \delta \Psi_v ||\hat f_1|| \Psi_w \rangle \, .
\label{eq:deltapsi}
\end{equation}
The correction $\delta \Psi_v$ is found by solving the system of
linear inhomogeneous equations
\begin{equation}
  (\hat H^{\rm eff} - E_v )\delta \Psi_v = - (\hat f_2+\delta V^{N-2}) \Psi_v.
\label{eq:DL}
\end{equation}
Here $\hat f_1$ and $\hat f_2$ are electric dipole and PNC interaction
operators ($\hat f_1 = {\bm d}, \hat f_2 = H_{\rm PNC}$ or
vice versa). 

\subsection{Accuracy of the calculations}

\begin{table}
\caption{Magnetic dipole hyperfine structure constants $A$ (MHz) for the
  $^3$P$^o_1$ and $^1$P$^o_1$ states of $^{171}$Yb, comparison with experiment.} 
\label{t:hfs}
\begin{ruledtabular}
\begin{tabular}{l rr}
 State & Calculations & Experiment\footnotemark[1] \\
\hline
 $^3$P$^o_1$ & 4460 &  3958 \\
 $^1$P$^o_1$ & -819 & -1094 \\
\end{tabular}
\footnotemark[1]{Reference~\cite{Ybhfs}.}
\end{ruledtabular}
\end{table}

Accuracy of very similar calculations of polarizabilities of ytterbium
were studied in detail in our previous work~\cite{DD10} and were found
to be about 5\%. However, we cannot claim the same accuracy for
present calculations due to two important differences. First, there is
a resonance contribution to the PNC amplitude involving the
$^1$P$^o_1$ state. Energy interval between the $^3$D$_1$ and
$^1$P$^o_1$  states is very small. Its experimental values is just
579~cm$^{-1}$. The term in (\ref{eq:e2}) involving the $^1$P$^o_1$
state gives more than 80\% of the total PNC amplitude. Even very
accurate calculations may give significantly different value of small
energy interval which would lead to large error in the PNC
amplitude. One way around this problem is to separate the resonance
term from the rest of the sum and use the experimental energy for the
denominator. We use a technically more simple procedure. We have
rescaled the correlation operator $\hat \Sigma$ to fit the interval
exactly. As a result, the contribution of the error in the energy
denominator to the error in the amplitude is small.

Another important difference of present calculations from those of
Ref.~\cite{DD10} is that we need to calculate matrix elements of weak
interaction which are sensitive to the wave functions on short
distances. A way to test the wave functions on short distances is to
calculate hyperfine structure (hfs) constants.  

Calculated and experimental values of the magnetic dipole hyperfine
structure constants $A$ for the $^3$P$^o_1$ and $^1$P$^o_1$ states of
$^{171}$Yb are presented in Table~\ref{t:hfs}. The first calculated
hfs constant is larger than the experimental one by 13\%, the second
is smaller by 25\%. The reason for the calculated hyperfine constant
of the $^1$P$^o_1$ state to differ significantly from the  the
experimental 
value is the same as for the electric dipole transition amplitude
between this and ground state - the admixture of the $4f^{13}5d6s^2 \
(7/2,3/2)^o_1$ state at $E=28857 {\rm cm}^{-1}$ (see Ref. \cite{DD10}
for details). This admixture is small. However, it can change hfs of
the $^1$P$^o_1$ state significantly due to the large hfs in the admixed
state. In contrast, it cannot change that much the weak matrix element
between the $^1$P$^o_1$ and $4f^{14}5d6s \ ^3$D$_1$ states. This is because
the transition between the $4f^{13}5d6s^2$ and the $4f^{14}5d6s \
^3$D$_1$ states in zero approximation is the $ 6s \rightarrow 4f$ transition
 and corresponding weak matrix element is zero. Therefore, poor accuracy
 for the hfs of
the $^1$P$^o_1$ state is not a good indicator for the accuracy of the
PNC calculations. A 13\% error in the hfs of the $^3$P$^o_1$ state
gives a more realistic estimate for the uncertainty.

We stress that the uncertainty in the ratio of SD and SI PNC amplitudes
($E^{\prime\prime}/E^{\prime}$) is significantly lower. Tests show
that this ratio is three to five times less sensitive to the variation
of the calculation procedure than each of the amplitudes. We believe
that 10\%  is a reasonable estimate for the theoretical uncertainty
for this ratio.   

\section{Results}

\begin{table}
\caption{PNC amplitudes ($z$-components) for the $|6s^2,^1$S$_0,F_1
  \rangle \rightarrow  |6s5d,^3$D$_1,F_2\rangle$ transitions in
  $^{171}$Yb and $^{173}$Yb in units of $E^{\prime}Q_W$ and $10^{-9} iea_0$.} 
\label{t:1}
\begin{ruledtabular}
\begin{tabular}{cccc rr}
$A$ & $I$ & $F_1$ & $F_2$ & \multicolumn{2}{c}{PNC amplitude} \\
 &&&& \multicolumn{1}{c}{units: $E^{\prime}Q_W$} &
 \multicolumn{1}{c}{units: $10^{-9} iea_0$} \\ 
\hline
171  & 0.5 & 0.5 & 0.5 & $-(1/3)(1-0.0161\varkappa)$ &
                                $ 6.15(1-0.0161\varkappa)$ \\

     &     & 0.5 & 1.5 & $\sqrt{2/9}(1+0.0081\varkappa)$ &
                                $-8.70(1+0.0081\varkappa)$ \\
     &     &     &     &  \\
173  & 2.5 & 2.5 & 1.5 & $-\sqrt{4/45}(1-0.0111\varkappa)$ &
                                $5.61(1-0.0111\varkappa)$ \\

     &     & 2.5 & 2.5 & $-\sqrt{5/21}(1-0.0032\varkappa)$ &
                                $9.18(1-0.0032\varkappa)$ \\

     &     & 2.5 & 3.5 & $\sqrt{2/21}(1+0.0079\varkappa)$ &
                               $-5.81(1+0.0079\varkappa)$ \\
\end{tabular}
\end{ruledtabular}
\end{table}

Calculations give the following value of the spin-independent PNC
amplitude of the $^1$S$_0 \rightarrow ^3$D$_1$ transition in
ytterbium:
\begin{equation}
  E_z^{\rm SI-PNC} = 1.123 \times 10^{-11} Q_W iea_0.
\label{ezprime}
\end{equation}
This corresponds to the following value of the reduced matrix element
\begin{equation}
  E^{\prime} = 1.945 \times 10^{-11} iea_0.
\label{eprime}
\end{equation}
The electron ($F$-independent) part of the reduced matrix element of
the spin-dependent PNC amplitude is found to be
\begin{equation}
  E^{\prime\prime} = 3.648 \times 10^{-11} iea_0, \label{eprime2}
\end{equation}
The effect of different nuclear size for $^{171}$Yb and $^{173}$Yb is
only 0.1\% for both SI and SD PNC amplitudes. It is neglected in
(\ref{ezprime}), (\ref{eprime}) and (\ref{eprime2}).
We use Fermi-type distribution for nuclear density $\rho$ with nuclear
radius $R_N=6.35$~fm for $^{171}$Yb and $R_N=6.37$~fm for 
$^{173}$Yb~\cite{Angeli}. 

The ratios of the SD and SI PNC amplitudes are
\begin{eqnarray}
  E^{\prime\prime}/(E^{\prime}Q_W) = -0.0198(20) &{\rm for}&
  ^{171}{\rm Yb}, \label{eprime4} \\
  E^{\prime\prime}/(E^{\prime}Q_W) = -0.0194(20)  &{\rm for}&
  ^{173}{\rm Yb}.
\label{eprime5}
\end{eqnarray}
The difference in these values is due to different weak nuclear
charge $Q_W$ ($Q_W = -94.75$ for $^{171}$Yb and $Q_W =-96.72$ for
$^{173}$Yb). The difference is within numerical uncertainty.

The results for the specific PNC amplitudes between different hyperfine
structure states of $^{171}$Yb  and $^{173}$Yb  are presented in
Table~\ref{t:1}. These results are obtained by substituting
(\ref{eprime}) and (\ref{eprime2})  into (\ref{1-2-1-2}),
(\ref{1-2-3-2}), (\ref{5-2-3-2}), (\ref{5-2-5-2}) and
(\ref{5-2-7-2}). The numerical factors before 
$\varkappa$ are proportional to ($E^{\prime\prime}/E^{\prime}$). The
theoretical uncertainty for these factors is about 10\%.
The expressions from the table or equations (\ref{1-2-1-2}),
(\ref{1-2-3-2}), (\ref{5-2-3-2}), (\ref{5-2-5-2}),
(\ref{5-2-7-2}) together with (\ref{eprime}) and (\ref{eprime2})
can be used to extract the value of
$\varkappa$ from the measurements. For example, for $^{171}$Yb
Eq.~(\ref{eq:kappa}) becomes
\begin{equation}
  \varkappa = 88(9)\frac{1+\sqrt{2}R}{(\sqrt{2}-R)}.
\label{eq:kappa1}
\end{equation}

\subsection{Comparison with other calculations}

Calculations of the spin-independent part of the PNC amplitude for
ytterbium were performed before in
Refs.~\cite{DeMille,Porsev95,Das97}. The spin-dependent amplitudes
were calculated before in Refs.~\cite{Singh99,Porsev00}. It is
convenient to compare the results in terms of $E^{\prime}$ (\ref{eq:si})
and $E^{\prime\prime}$ (\ref{eq:epp}) since these values are the same
for all hfs transitions. Refs.~\cite{DeMille,Porsev95,Das97} present
the values of the $z$-component of the SI PNC
amplitude. Refs.~\cite{Singh99,Porsev00} present reduced matrix
elements of the SD PNC interaction for each hfs
transition. Corresponding values of $E^{\prime}$ and
$E^{\prime\prime}$ can be easily extracted from this data using
formulas of present paper. The absolute values of the amplitudes
 are presented in
Table~\ref{t:2}. We have excellent agreement for $E^{\prime}$ with
DeMille~\cite{DeMille} and Porsev {\em et al}~\cite{Porsev95} while
the result of Das~\cite{Das97} is about 30\% smaller. We have good
agreement with both Porsev {\em et al}~\cite{Porsev00} and Singh
and Das~\cite{Singh99} for $E^{\prime\prime}$. The difference
between Ref.~\cite{Porsev00} and our result is 12\% which is within
our uncertainty. The difference between our
results for $E^{\prime\prime}$ and those of Ref.~\cite{Singh99} is
even smaller. But this is probably accidental. Note however that we
agree with Singh and Das~\cite{Singh99} on small, practically
negligible change of $E^{\prime\prime}$ from $^{171}$Yb to $^{173}$Yb
while Porsev {\em at al}~\cite{Porsev00} report a 3\% increase.
Such increase has no physical explanation and must be a numerical
effect. In our experience such effect can be a result of just one RPA
iteration after a change of nuclear radius from $^{171}$Yb to
$^{173}$Yb. Further iterations kill the difference. However, it is up
to the authors of \cite{Porsev00} to explain their results.

Note again that in Table ~\ref{t:2} we present only the absolute
values of the amplitudes, 
ignoring their signs. This is because the sign of an amplitude is not
fixed and has no physical meaning. However, the relative sign of the SI
and SD PNC amplitudes is not arbitrary. The SD dependent part of the PNC
amplitude must either increase or decrease the transition amplitude
depending on the sign of $\varkappa$. It is important to know the
relative sign of the amplitudes to be able to extract the sign of
$\varkappa$ from the measurements. To fix the relative sign of the two
PNC amplitudes one should calculate them 
using the same wave functions. This is how it is done in present work (see
Table~\ref{t:1} and formulas (\ref{1-2-1-2}),
(\ref{1-2-3-2}), (\ref{5-2-3-2}), (\ref{5-2-5-2}),
(\ref{5-2-7-2})). Another important advantage of the  simultaneous
calculation of both amplitudes is that $\varkappa$ can be expressed
via the ratio of the amplitudes. This ratio has much smaller
theoretical uncertainty than each of the amplitudes (see previous
section for discussion). 

Unfortunately, both previous calculations of the SD PNC amplitude in
Yb~\cite{Singh99,Porsev00} do not compare their results with the earlier
calculations of the SI PNC amplitudes~\cite{Porsev95,Das97} which
could be performed using different wave functions. This leads to the
uncertainty of the relative signs and larger errors in the ratios 
of the SI and SD amplitudes which are needed to extract the value of
 $\varkappa$ from the measurements.

\begin{table}
\caption{Spin-independent ($E^{\prime}$) and spin-dependent
  ($E^{\prime\prime}$) parts of the PNC amplitude (reduced matrix elements) for
  the $|6s^2,^1$S$_0 \rangle \rightarrow  |6s5d,^3$D$_1\rangle$ transition in
  ytterbium, comparison with other calculations. The signs of the
  amplitudes are omitted.} 
\label{t:2}
\begin{ruledtabular}
\begin{tabular}{llll}
& \multicolumn{1}{c}{$E^{\prime}$} & 
\multicolumn{2}{c}{$E^{\prime\prime}\ (10^{-11} iea_0)$} \\
& \multicolumn{1}{c}{$10^{-11} iea_0$} 
& \multicolumn{1}{c}{$^{171}$Yb} 
& \multicolumn{1}{c}{$^{173}$Yb} \\ 
\hline
DeMille~\cite{DeMille} & 1.9 & & \\
Porsev {\em et al}  &  1.97\footnotemark[1] &  4.13\footnotemark[2] &
                                              4.27\footnotemark[2] \\
Singh and Das &  1.33\footnotemark[3] &  3.68\footnotemark[4] &
                                        3.67\footnotemark[4] \\
This work     &  1.95            &   3.65  &  3.64            \\
\end{tabular}
\footnotemark[1]{Ref.~\cite{Porsev95},}
\footnotemark[2]{Ref.~\cite{Porsev00},}
\footnotemark[3]{Ref.~\cite{Das97},}
\footnotemark[4]{Ref.~\cite{Singh99}.}
\end{ruledtabular}
\end{table}

\subsection{Comparison with experiment}

The result of the measurement of the PNC amplitude of the $^1$S$_0
\rightarrow ^3$D$_1$ transition in $^{174}$Yb reported in~\cite{BudkerYbPNC}
reads 
\begin{equation}
|E^{\rm PNC}| = 8.7(1.4) \times 10^{-10} ea_0.
\label{eq:exp}
\end{equation}
Assuming a 13\% theoretical uncertainty and substituting weak nuclear
charge $Q_W=-97.71$ we get from (\ref{ezprime}) the following
theoretical value for the amplitude 
\begin{equation}
|E^{\rm PNC}| = 11.0(1.4) \times 10^{-10} ea_0.
\label{eq:theor}
\end{equation}
The values of (\ref{eq:exp}) and (\ref{eq:theor}) agree within the
declared uncertainty. 

To measure the constant of spin-dependent PNC interaction
($\varkappa$) more accurate measurements are needed for $^{171}$Yb or
$^{173}$Yb. The work is in progress at Berkeley~\cite{BudkerYbPNC}.

\section{Conclusion}

We present simultaneous calculation of the spin-independent and
spin-dependent PNC amplitudes of the $6s^2 \ ^1$S$_0 \rightarrow 6s5d
\ ^3$D$_1$ transition in ytterbium. The results are to be used for
accurate interpretation of future measurements in terms of the
parameter of the spin-dependent PNC interaction $\varkappa$. Both,
sign and value of $\varkappa$ can be determined. Theoretical
uncertainty is at the level of 10\%.

\acknowledgements

The authors are grateful to M. G. Kozlov and S. G. Porsev for useful
discussions. 
The work was supported in part by the Australian Research Council.

\end{document}